\documentclass{aa}
\usepackage[normalem]{ulem}
\usepackage[dvipsnames]{xcolor}
\usepackage{tikz}
\usetikzlibrary{tikzmark}
\usetikzlibrary{calc}
\usetikzlibrary{shapes.geometric, arrows}
\usepackage{color,graphicx}
\usepackage{amsmath} 
\usepackage{natbib}
\usepackage[varg]{txfonts} 
\bibpunct[]{(}{)}{;}{a}{}{,} 

\newcommand{\vlos}{V_{\rm los}}
\newcommand{\vzon}{V_{\rm zonal}}
\newcommand{\vband}{\vzon^{\rm (band)}}
\newcommand{\vb}{V_{\rm background}}
\newcommand{\HL}{{HL1}}
   
\begin{document}

\title{Doppler velocity of $m=1$ high-latitude inertial mode\\ over the last five sunspot cycles}
\titlerunning{High-latitude inertial waves}

\author{
Zhi-Chao Liang\inst{\ref{mps}} and Laurent Gizon\inst{\ref{mps},\ref{gottingen},\ref{NYUAD}}
}

\institute{
Max-Planck-Institut f\"ur Sonnensystemforschung, Justus-von-Liebig-Weg 3, 37077 G\"ottingen, Germany\\
\email{zhichao@mps.mpg.de; gizon@mps.mpg.de} \label{mps}
\and Institut f\"ur Astrophysik und Geophysik, Georg-August-Universit\"at G\"ottingen,  37077 G\"ottingen, Germany \label{gottingen}
\and Center for Astrophysics and Space Science, NYUAD Institute, New York University Abu Dhabi, Abu Dhabi, UAE \label{NYUAD}
}

\date{Received $\langle$date$\rangle$ / Accepted $\langle$date$\rangle$}

\abstract
{ 
A variety of solar global modes of oscillation in the inertial frequency range have been identified in maps of horizontal flows derived from GONG and HMI data. Among these, the high-latitude mode with azimuthal order $m=1$ ($\HL$) has the largest amplitude and plays a role in shaping the Sun's differential rotation profile.
} { 
We aim to study the evolution of the $\HL$ mode parameters,  utilizing Dopplergrams from the Mount Wilson Observatory (MWO), GONG, and HMI, covering together five solar cycles  since 1967.
} { 
We calculated the averages of line-of-sight Doppler signals over longitude, weighted by the sine of longitude with respect to the central meridian, as a proxy for zonal velocity at the surface.
We measured the mode's power and frequency from these zonal velocities at high latitudes in sliding time windows of three years.
} { 
The $\HL$ mode is easily observed in the  maps of zonal velocity at latitudes above 50 degrees, especially during solar minima.
The mode parameters measured from the three independent data sets  are consistent during their overlapping periods and agree with previous findings using HMI ring-diagram analysis.
We find that the amplitude of the mode undergoes very large variations, taking maximum values at the start of solar cycles 21, 22 and 25, and during the rising phases of cycles 23 and 24.
The mode amplitude is anticorrelated with the sunspot number ($\textrm{corr} =-0.50$) but not correlated with the polar field strength. 
Over the period 1983--2022 the mode amplitude is strongly anticorrelated with the rotation rate at latitude $60^\circ$ ($\textrm{corr}=-0.82$), i.e., with the rotation rate near the mode's critical latitude. 
The mode frequency variations are small and display no clear solar cycle periodicity above the noise level ($\sim \pm 3$~nHz).
Since about 1990, the mode frequency follows an overall  decrease of $\sim 0.25$ nHz/year, consistent with the long-term decrease of the angular velocity at $60^\circ$ latitude.
} { 
We have shown that the amplitude and frequency of the $\HL$ mode can be  measured over the last five solar cycles, together with the line-of-sight projection of the velocity eigenfunction.
We expect that these very long time series of the mode properties will be key to understand the dynamical interactions between the high-latitude modes, differential rotation, and (possibly) magnetic activity.
}


\maketitle

\section{Introduction}
Rossby and inertial modes of solar oscillation are observed in long-duration  time series of the horizontal velocity field on/near the solar surface  \citep{Loeptien2018,Gizon2021, Hanson2022}. These horizontal flows may be inferred at a daily cadence using local helioseismology applied to the Global Oscillation Network Group \citep[GONG:][]{Harvey1996} and the Helioseismic and Magnetic Imager onboard the Solar Dynamics Observatory \citep[SDO/HMI:][]{Scherrer2012} data sets, or using local correlation tracking of granulation (seen in SDO/HMI intensity images). The observed modes have periods of order the solar rotation period and their surface eigenfunctions are  quasi-toroidal.  Linear eigenvalue solvers enable us to identify  the various modes that are clearly resolved in frequency space and to infer their radial eigenfunctions throughout the convection zone \citep{Bekki2022a}. The equatorial Rossby modes have maximum colatitudinal velocity, $v_\theta$, at the equator, while the high-latitude modes have maximum longitudinal velocity, $v_\phi$, in the polar regions. Nearly all modes have critical latitudes where their phase speeds equal the local  differential rotation velocity. In the observations, the inertial modes are unambiguously identified as global modes because each mode oscillates at the same frequency over a large range of  latitudes \citep{Gizon2021}. \cite{Waidele2023} and \cite{Lekshmi2024} measured the temporal variations of the equatorial Rossby mode properties using the horizontal flows from helioseismic analysis of HMI and GONG data, and reported that the mode power (frequency shift) averaged over azimuthal wavenumber $m$ is correlated (anticorrelated) with the sunspot cycle.
We refer the reader to \citet{Gizon2024} for a review on the topic of solar inertial modes.

The high-latitude inertial mode with $m=1$ (hereafter $\HL$ for short) is of particular interest. It has the largest amplitude among all  modes in the inertial frequency range \citep[time-averaged $v_\phi\sim 10$~m/s, see][]{Gizon2021} and it is driven by a baroclinic instability \citep{Bekki2022a}. This mode is very sensitive to the latitudinal entropy gradient and to other important properties of the solar convection zone, such as the superadiabatic temperature gradient \citep{Gizon2021, Bekki2022a, Bekki2024}. The mode is anti-symmetric across the equator in $v_\phi$ and symmetric in $v_\theta$. Its mean frequency over the period 2010\,--\,2020 is measured to be $\nu_\HL^{\rm Carr} = -86.3\pm 1.6$~nHz in the Carrington frame, where the negative sign indicates retrograde propagation.
In an inertial frame, the sidereal frequency is  
\begin{equation} \label{eq:nu_sidereal}
    \nu_\HL = -86.3 + 456.0~\textrm{nHz} = 369.7~\textrm{nHz}
\end{equation} 
and in the Earth frame the  synodic frequency is 
\begin{equation} \label{eq:nu_obs}
    \nu_\HL^{\rm synodic} = 369.7-31.7~\textrm{nHz} = 338.0~\textrm{nHz} .
\end{equation}
Above $50^\circ$ latitude, the mode has a characteristic spiral pattern in $v_\phi$, which, according to models, depends sensitively on the properties of the convection zone  \citep{Gizon2021}. The flow pattern associated with the  $m=1$ mode was first reported by \cite{Hathaway2013} using a supergranulation tracking method, although it was then misidentified  as giant cell convection sheared by differential rotation. \citet{Bogart2015} later confirmed the existence of the $m=1$ polar flow pattern using ring-diagram analysis.

It came to our attention  only recently that the  $m=1$ high-latitude inertial mode of oscillation was clearly seen above $50^\circ$ latitude  in the  ``zonal velocity'' time-series  computed from Mt.~Wilson Observatory (MWO) Dopplergrams by 
\citet[][, his figure~5, top panel]{Ulrich1993} and
\citet[][, his figure~1, top panel]{Ulrich2001}.
\citet{Ulrich1988} defined  the zonal velocity as
\begin{equation} \label{eq:vzon}
    \vzon (\theta) = \frac{\sum_{|L|<30\degr} (\sin L)    [\vlos(\theta,L) - \vb(\theta,L)]}{\sum_{|L|<30\degr} |\sin L|} ,
\end{equation}
where $\vlos$ is a Dopplergram and   $\vb$ is the Doppler image consisting of the limb shift plus the line-of-sight (LOS) projection of the Sun's axisymmetric flows (rotation and meridional circulation). The co-latitude is denoted by $\theta$ and  $L=\phi-\phi_{\rm c}$ is the heliographic longitude measured with respect to  the central meridian $\phi_{\rm c}$. The zonal velocity  informs us about the slowly-varying large-scale components of the  longitudinal velocity $v_\phi$ on the Sun. The velocity $\vzon$ can be used to measure  the `torsional oscillation', and also to detect the inertial modes.

In figure~4 of \citet{Ulrich2001}, the $m=1$ high-latitude mode is visible in $\vzon$ at latitudes $\pm 58.5^\circ$ as narrow peaks of power with very high signal-to-noise ratio at a sidereal frequency close to $0.8 \times (\textrm{Carrington Period})^{-1} = 0.8\times 456~ \textrm{nHz} = 365$~nHz. This frequency corresponds to $\nu_\HL$ given in Eq.~\eqref{eq:nu_sidereal} and is slightly above the local rotation rate. At the lower latitudes (below $40^\circ$), other peaks of excess power are seen at frequencies that follow the local frequency of the surface differential rotation (increases toward the equator); these peaks are unrelated to the $\HL$ mode, but are likely due to  velocity features that are associated with magnetic activity  advected at the local rotation rate.

In the present paper, we use all three available series of full-disk Dopplergrams (MWO, GONG and HMI) to study the long-term  evolution of the $\HL$ mode. We find that the mode is visible in $\vzon$  in the latitude range $60^\circ$\,--\,$75^\circ$ throughout the entire period from 1967 to 2024.  We characterize the evolution of the $\HL$ mode parameters and study their correlations with the sunspot number, the polar field strength, and the solar rotation rate. The steps to obtain the $\HL$ mode parameters are detailed in Sections 2, 3, and 5, and summarized in Fig.~\ref{flowchart}. We note that preliminary results using only GONG and HMI were presented  by \cite{Gizon2024}.

\begin{figure}
\centering

\begin{tikzpicture}[node distance=1.5cm and 2cm]

\tikzset{
    *|/.style={
        to path={
            (perpendicular cs: horizontal line through={(\tikztostart)},
                                 vertical line through={(\tikztotarget)})
            -- (\tikztotarget) \tikztonodes
        }
    }
}


\tikzstyle{startstop} = [rectangle,  minimum width=1cm, minimum height=0.8cm, text centered, draw=black]

\tikzstyle{process} = [rectangle, minimum width=3cm, minimum height=0.8cm, text centered, draw=black, fill=gray!10]

\tikzstyle{process2} = [rectangle, minimum width=1cm, minimum height=0.6cm, text centered, draw=black, fill=gray!10]

\tikzstyle{arrow} = [thick, ->, >=stealth, line width=0.3mm]


\node[align=center] (GONG) [startstop] {{\color{Green}GONG}\\Dopplergrams\\from \texttt{tdvzi}\\(unmerged)};

\node[align=center] (MWO) [startstop, left of=GONG, xshift=-1cm] {{\color{orange}MWO}\\Dopplergrams};

\node[align=center] (HMI) [startstop, right of=GONG, xshift=1.3cm] {{\color{red}HMI}\\Dopplergrams\\from \texttt{hmi.v\_720s}};

\node[align=center] (ulrich2001) [process, below of=MWO, yshift=-0.5cm] {Static $V_{\rm background}$\\ removed by\\ \citet{Ulrich2001}};

\node[align=center] (step1) [process, below of=GONG, yshift=-2.2cm] {Reduce spatial resolution to $10$~arcsec/pixel\\at a daily cadence};

\node[align=center] (step2) [process2, below of=step1, xshift=1.8cm] {Subtract $V_{\text{background}}$ (Sect.~\ref{sec:Vbkg})};

\node (step3) [process2, below of=step2, xshift=0.95cm] {Subtract fringes};

\node (step4) [process, below of=step3, xshift=-2.75cm] {Compute $\vzon$ (Eq.~\eqref{eq:vzon}) and filter ($200$~nHz high-pass)};

\node (step5) [process, below of=step4] {Compute north-south antisymmetric component $\vzon^{(-)}$};

\node (step6) [process, below of=step5] {Compute $\vband$ by averaging $\vzon^{(-)}$ over $60^\circ$--$75^\circ$  (Eq.~\eqref{eq:Vband})};

\node (step7) [process, below of=step6] {Compute power spectra of $\vband$ for 3-year chunks};

\node[align=center] (step8) [process, below of=step7] {Fit Lorentzian profile to $\HL$ peak near 338 nHz\\to measure mode parameters (amplitude and frequency)};

\node[align=center] (step9) [process, below of=step8] {Scale up LOS mode amplitude by a factor of $2.8$\\to estimate the maximum horizontal velocity};


\draw [arrow] (MWO) -- (ulrich2001);
\draw [arrow] (ulrich2001.south) -- ([xshift=-2.5cm]step1.north);
\draw [arrow] (GONG) -- (step1);
\draw [arrow] (HMI.south) -- ([xshift=2.8cm]step1.north);

\draw [arrow,*|] (step1.south) to (step2.north);
\node[Green] at (1.1cm,-4.5cm) {GONG};
\node[red] at (2.3cm,-4.5cm) {HMI};

\draw [arrow,*|] (step2.south) to (step3.north);
\node[red] at (3.3cm,-5.9cm) {HMI};

\draw [arrow] ([xshift=-1.5cm]step1.south) -- ([xshift=-1.5cm]step4.north);
\node[orange] at (-2.1cm,-5.9cm) {MWO};

\draw [arrow] ([xshift=-1.1cm]step2.south) -- ([xshift=0.7cm]step4.north);
\node[Green] at (0.1cm,-6.7cm) {GONG};

\draw [arrow] (step3.south) -- ([xshift=2.75cm]step4.north);

\draw [arrow] (step4) -- (step5);
\draw [arrow] (step5) -- (step6);
\draw [arrow] (step6) -- (step7);
\draw [arrow] (step7) -- (step8);
\draw [arrow] (step8) -- (step9);
\end{tikzpicture}
\caption{ \label{flowchart}
Schematics of data processing steps involved in the measurement of the velocity amplitude of the $\HL$ mode from MWO, GONG and HMI time series of Dopplergrams.}
\end{figure}

\section{Detection of $\HL$ mode in  MWO, GONG and HMI Dopplergrams}

The time periods used in the analysis are from 1967 to 2012 for MWO data, from 2001 to 2024 for GONG data (\texttt{tdvzi}), and from 2010 to 2024 for HMI data (\texttt{hmi.v\_720s}). The original time cadences and spatial resolutions of the data sets are quite different. For the purpose of this work, we take one frame per day and reduce the resolution to 10\arcsec per pixel by binning.

\subsection{Correction of GONG and HMI Dopplergrams} \label{sec:Vbkg}

The static limb shift, differential rotation, and meridional circulation velocity signals were already subtracted from the MWO data by \citet{Ulrich2001}, thus no $\vb$ subtraction is necessary for this dataset.
For the other two datasets, we perform low-order polynomial fits to $\vlos$ to obtain $\vb$.
For the limb shift, we fit a model that consists of a polynomial of order five in $\mu$, where $\mu$ is the cosine of the center-to-limb angle.
For the LOS projection of the trends in the east-west and north-south directions, we fit the following models:
\begin{align}
& \vlos^{\rm (EW)} (\theta, L) =  \cos B_0 \sin\theta  \sin L   \sum_{k=0}^{4} b_k (\cos\theta)^k ,\\
& \vlos^{\rm (NS)} (\theta, L) =  (\sin B_0 \sin\theta - \cos B_0 \cos\theta\cos L)  \sum_{l=1}^{4} c_l P_l^1 (\cos\theta) ,
\end{align}
where $B_0$ is the heliographic latitude at the disk center and the $P_l^1$ are associated Legendre polynomials of degree $l$ and order one. The model parameters are estimated using a least-squares method.
We correct the $\vlos$ by subtracting daily fits to the limb shift and $\vlos^{\rm (NS)}$. For $\vlos^{\rm (EW)}$, however, we subtract a long-term average instead, in order to preserve the zonal velocity component of low-$m$ inertial modes in $\vlos$.
We note that we apply the procedure to the unmerged GONG data (\texttt{tdvzi}), because the merged GONG data (\texttt{mrvzi}) are affected by the removal of a surface fit to the data, which undesirably reduces the amplitude of the low-$m$ modes.

\begin{figure}
    \centering
    \includegraphics[width=\hsize]{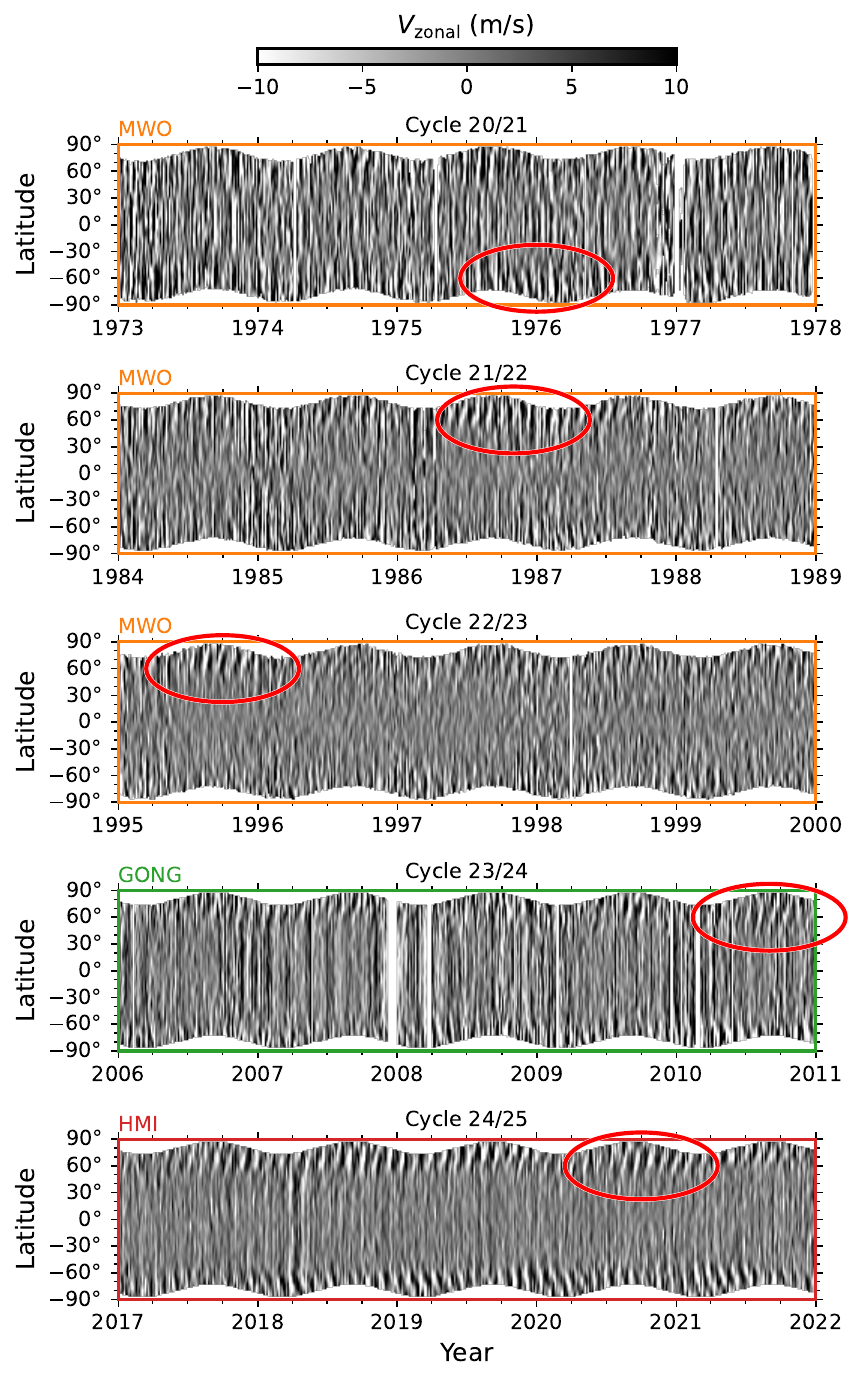}
    \caption{
    Example supersynoptic maps of $\vzon$ in the Earth frame computed from the MWO (top three panels), GONG (fourth panel) and HMI data (bottom panel) near cycle minima. The full data set is shown in Fig.~\ref{fig:synoptic-all}).
    The $\HL$ mode may be seen from time to time as stripes in $\vzon$ above $50\degr$ latitude (see regions in red ellipses).
    For clarity, the maps were smoothed in latitude with a Gaussian kernel of width $10\degr$.
    }
    \label{fig:synoptic}
\end{figure}

The HMI  Doppler time series also  contains a number of jumps, which coincide with the re-tuning  of the optical filter system \citep{Couvidat2016,Hoeksema2018}.
This affects the interference fringes. In the \texttt{hmi.v\_720s} data series, the fringes were largely removed from 1 October 2012 onward \citep[see][, their section~2.8.2 and their figures~9 and 10]{Couvidat2016}. However the fine structure of the fringes still remains in this data set (see Fig.~\ref{fig:fringes}).
To remove the HMI fringes, we subtract the median values of the images computed over time intervals between adjacent re-tuning dates.

\subsection{Computation of zonal velocity}

In order to compute the zonal velocity for MWO, GONG and HMI using  Eq.~\eqref{eq:vzon}, 
we remap the corrected Dopplergrams  onto heliographic coordinates using the equidistant cylindrical projection with a map scale of $0.6^\circ$ per pixel, to obtain $(\vlos-\vb)$ as a function $\theta$, $L$, and time. 
We choose to sum  over longitudes $|L|<30^\circ$ and average over nine consecutive days, in order to  follow the procedure described by \cite{Ulrich1988}. The resulting time series of zonal velocity, $\vzon(\theta,t)$, are kept in the Earth frame.

\begin{figure}
    \centering
    \includegraphics[width=\hsize]{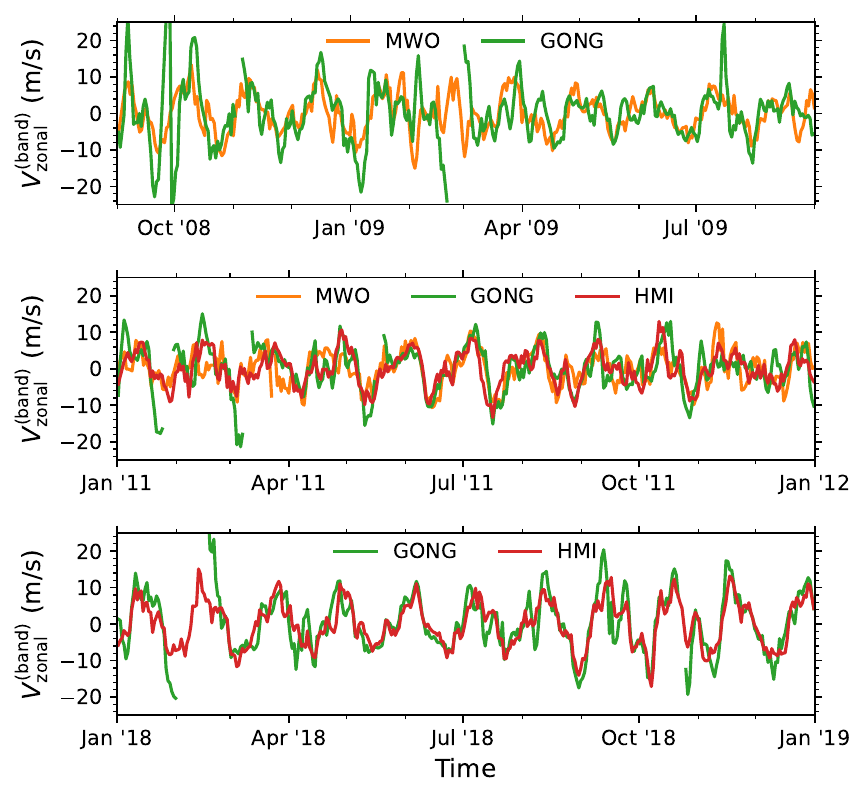}
    \caption{ \label{fig:Vband}
   $\vband(t)$  during different overlapping periods of the  MWO, GONG and HMI data. The period of the predominant oscillations seen in $\vband$ is about 34~days, which corresponds to the $\HL$ mode frequency near $\nu_\HL^{\rm synodic}=338$~nHz.
    }
\end{figure}

\subsection{Visual detection of $\HL$ mode in supersynoptic maps}

To enhance the visibility of the $\HL$ mode in $\vzon$, we apply a high-pass temporal filter to remove the slow variations below 200~nHz, including the `torsional oscillation' and the $B_0$-angle periodicities.
Figure~\ref{fig:synoptic-all} shows all the supersynoptic maps of $\vzon$ as functions of time and latitude for the three data sets (MWO, GONG, HMI).  The black and white stripes at high latitudes correspond to the $\HL$ inertial mode. Importantly, these stripes are consistent between the overlapping data sets (see arrows in Fig.~\ref{fig:synoptic-all}).
Figure~\ref{fig:synoptic} shows selected time periods from Fig.~\ref{fig:synoptic-all} showing only five years of data around each solar cycle minimum. The mode amplitude is strongest during these quiet Sun periods.

\subsection{Consistency between datasets during overlapping time periods}

Since the global mode $\HL$ is antisymmetric across the equator in $\vzon$, we compute the north-south antisymmetric zonal velocity $\vzon^{(-)}(\theta,t)=[\vzon(\theta,t) - \vzon(\pi-\theta,t)]/2$ to reduce noise.
We then compute an average of $\vzon^{(-)}$ over the latitudinal range $60\degr$\,--\,$75\degr$ where the mode is easily detectable:
\begin{equation} 
\label{eq:Vband}
    \vband(t) = \left\langle \vzon^{(-)}(\theta,t)\right\rangle_{|\theta-22.5^\circ|\leq 7.5^\circ} .
\end{equation}
Figure~\ref{fig:Vband} shows extracts from $\vband$ for overlapping time segments between MWO, GONG, and HMI. The consistency between the overlapping data sets is clear.

\section{Power spectra of zonal velocity}
\label{sect:fullres}

We then Fourier transform each of the MWO, GONG, and HMI $\vband(t)$ to obtain three power spectra,
\begin{equation} \label{eq:power}
    P(\nu) = \frac{1}{\Delta\nu\ N_t^2\ \eta}\left|\sum_t \vband(t)\ e^{-2\pi\textrm{i} \nu t} \right|^2 , 
\end{equation}
where  $N_t=16824$ for MWO,  $N_t=8462$ for GONG,  $N_t=5267$ for HMI, with a common temporal sampling $\Delta t = 1$~day. The quantity $\Delta \nu = 1 / (N_t \Delta t)$ is the frequency resolution, equal to $0.7$~nHz for MWO, $1.4$~nHz for GONG, and $2.2$~nHz for HMI.
To correct for missing data, we divide the power by the duty cycle $\eta=0.96$ for MWO, $\eta=0.96$ for GONG, and $\eta=1$ for HMI.

\begin{figure}
    \centering
    \includegraphics[width=0.9\hsize]{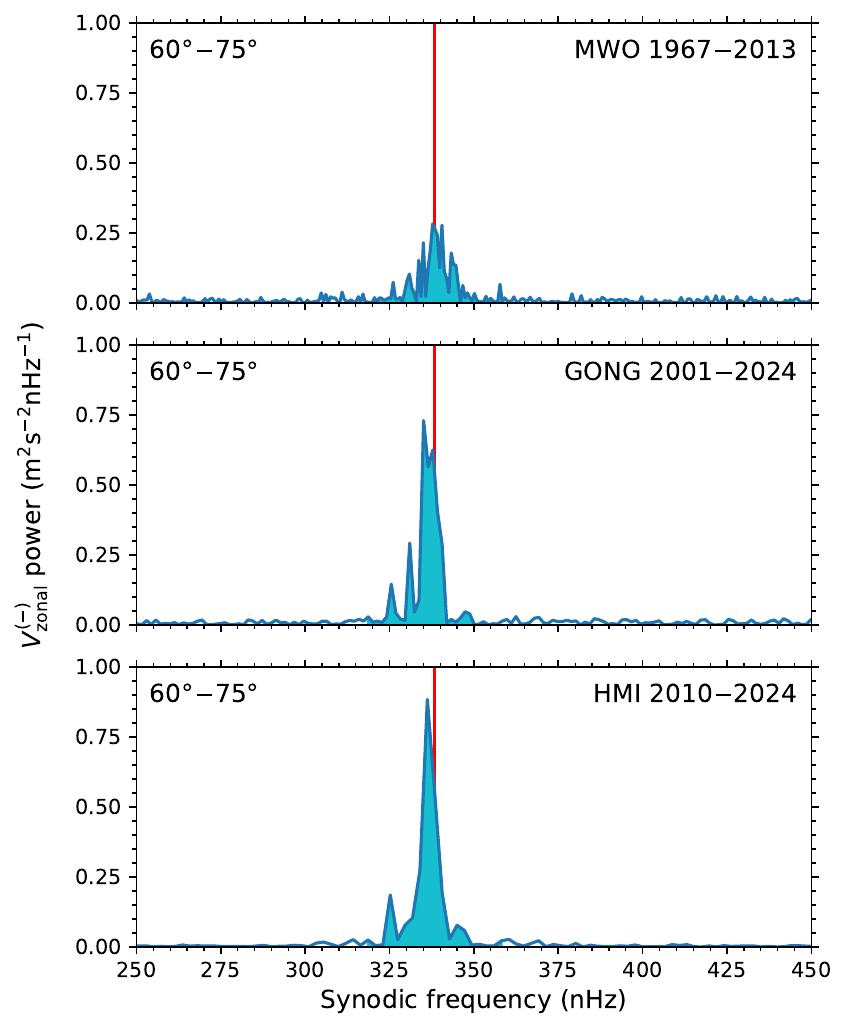}
    \caption{ \label{fig:fullres}
    Power spectra of $\vband$ in the frequency range 250--450 nHz (in the Earth frame),  computed from the  MWO, GONG and HMI Dopplergrams (from top to bottom respectively).
    The vertical red lines at $338$~nHz mark the reference   synodic frequency of the $\HL$ mode.
    }
\end{figure}

Figure~\ref{fig:fullres} shows $P(\nu)$ for the full time series of the three data sets.
In all cases, excess power is detected around the reference $\HL$ mode frequency $\nu_\HL^{\rm synodic}=338$~nHz.
The mode power differs between the three data sets.
This is not because of varying data quality, but because they do not cover the same time periods.
We shall confirm later that the mode power and frequencies measured from the three data sets are consistent during their overlap periods.
The central frequency of the peak in the HMI $P(\nu)$ is 2~nHz off from the reference $\nu_\HL^{\rm synodic}$, although the reference frequency was obtained using the HMI data in a similar time period.
One possible explanation is that \cite{Gizon2021} obtained the $\HL$ mode frequency using the latitude range $37.5\degr$\,--\,$67.5\degr$ where the excess power around the local differential rotation rate (highlighted by the blue curves in their figure~1) remains significant and thus the measured frequency was slightly biased.

In all three data sets the background noise level is extremely low. 
The full width of the peak is approximately $w$=4$\sim$6~nHz. This corresponds to an e-folding lifetime of $1/(\pi w) \sim 2$~years.

As will be seen later in the three-year power spectra of Fig.~\ref{fig:pow-vs-nu} (middle column), the mode is continuously visible since 1969.

\section{Velocity pattern associated with the $\HL$\ mode}

Let us infer the LOS velocity perturbation associated with the $\HL$ mode from the series of Dopplergrams.
We align the disk centers of the HMI $\vlos$ images, rotate the images by the nominal roll angle provided by the HMI team ($-$\texttt{CROTA2}), and re-scale the disk radius to a common value.
Instead of taking one frame per day, we compute daily averages of the images to reduce the noise.
At each pixel $(x,y)$ on the disk, we apply a band-pass filter to the data in the Fourier domain centered around the synodic mode frequency to obtain
\begin{equation} \label{eq:filt}
    \vlos^{\HL} (x,y,t):= \frac{1}{N_t}\sum_\nu \mathrm{filt}\left(\frac{\nu - 338\ \textrm{nHz}}{75\ \textrm{nHz}}\right) \hat{V}_\textrm{los}(x,y,\nu)\ e^{2\pi\textrm{i} \nu t},
\end{equation}
where  filt($\xi$) is a symmetric function equal to 1 for $|\xi|<1/3$, tapered to zero with raised cosines, and with a full width at half maximum of 1. In the above expression, $\hat{V}_\textrm{los}$ denotes the temporal FFT of $\vlos$.
\begin{figure}
  \centering
  \includegraphics[width=0.8\hsize]{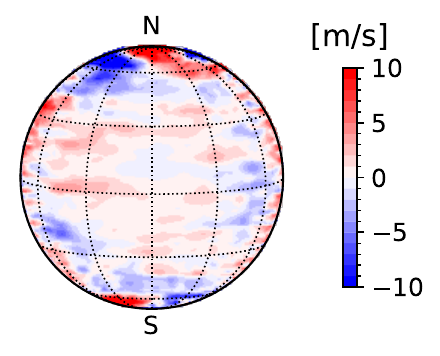}
  \caption{ \label{fig:vlos}
  Line-of-sight Doppler velocity associated with the $\HL$ mode as viewed from the Earth on 09 September 2018 when the solar tilt angle $B_0$ is $7.24\degr$, extracted from time series of daily HMI images by applying a band-pass filter around $\nu_\HL^{\rm synodic}$ (see Eq.~\eqref{eq:filt}).
  The LOS velocity pattern in the polar regions is associated with the $\HL$ mode.
  The image was smoothed using a 2D Gaussian kernel with a FWHM of $5\degr$.
  A comparison with the reconstructed $\vlos^\HL$ using $5\degr$-tile ring-diagram flow maps is shown in Fig.~\ref{fig:cf-eigenfunc} and in the movie \citep{movie}.}
\end{figure}

Figure~\ref{fig:vlos} shows an example of a filtered Dopplergram during the solar minimum in 2018, at a time when the north pole is tilted toward the Earth. The image was smoothed using a 2D Gaussian kernel with a FWHM of $5\degr$ to reduce noise on small spatial scales.
The horizontal velocity of the $\HL$ mode projected onto the line of sight is clearly seen above $50\degr$ latitude in both hemispheres.
Notice that there appears to be a polar-crossing flow with an amplitude over 10~m/s associated with the $\HL$ mode \citep[in agreement with an early statement by][, page 35]{Ulrich1993}.
In the left column of Fig.~\ref{fig:cf-eigenfunc}, we can see the line-of-sight projection of the mode velocity at different times (different phases) with different $B_0$ angles.
The predominant pattern in $\vlos^\HL$ at high latitudes is mostly north-south antisymmetric.

\begin{figure*}[h]
\centering

  \begin{tabular}{ccc}
  \begin{minipage}[b]{64mm}
    \resizebox{\hsize}{!}{\includegraphics{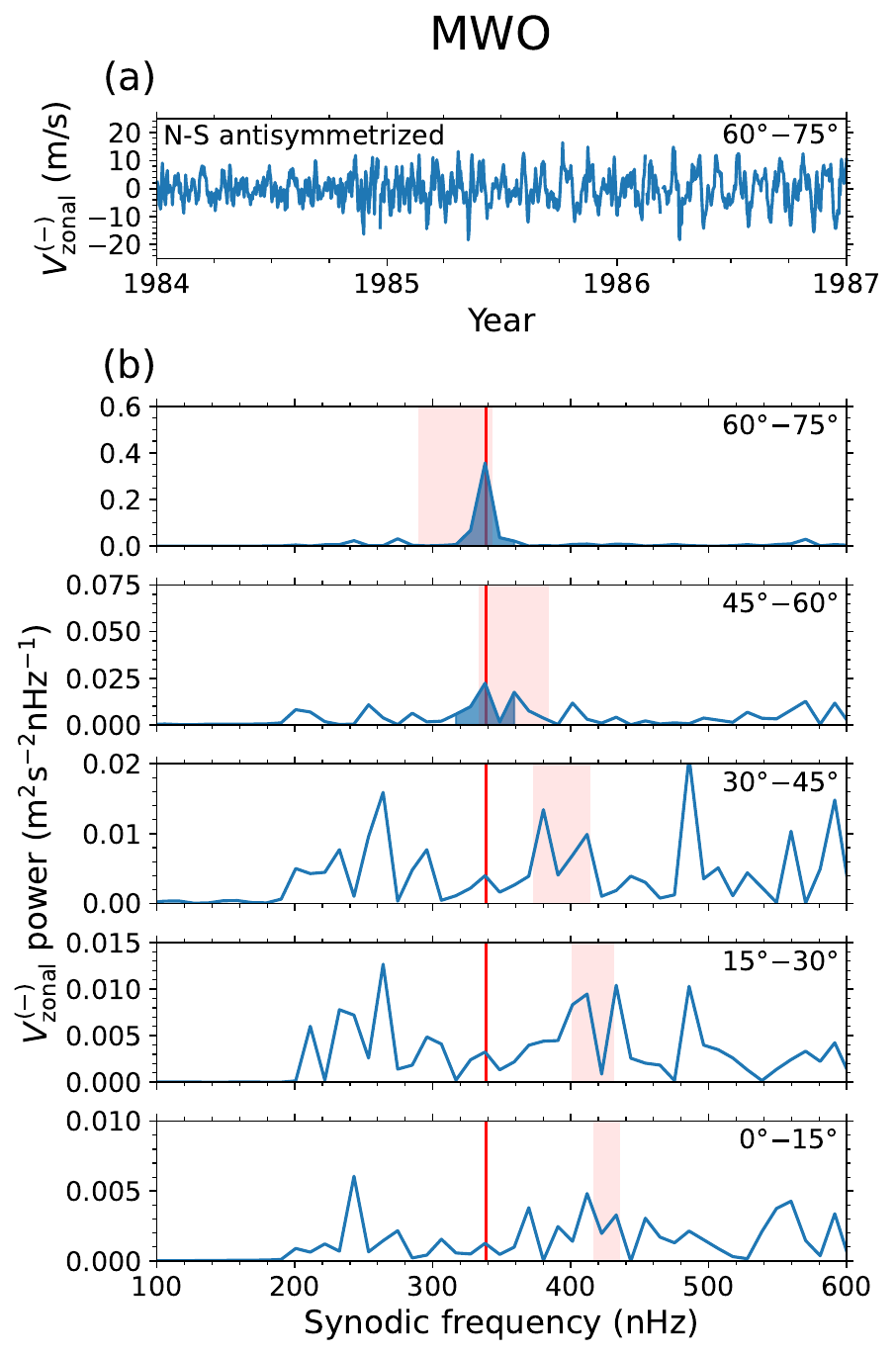}}
  \end{minipage}
  \tikzmark{end1}
  &
  \tikzmark{start1}
  \begin{minipage}[b]{34mm}
    \resizebox{\hsize}{!}{\includegraphics{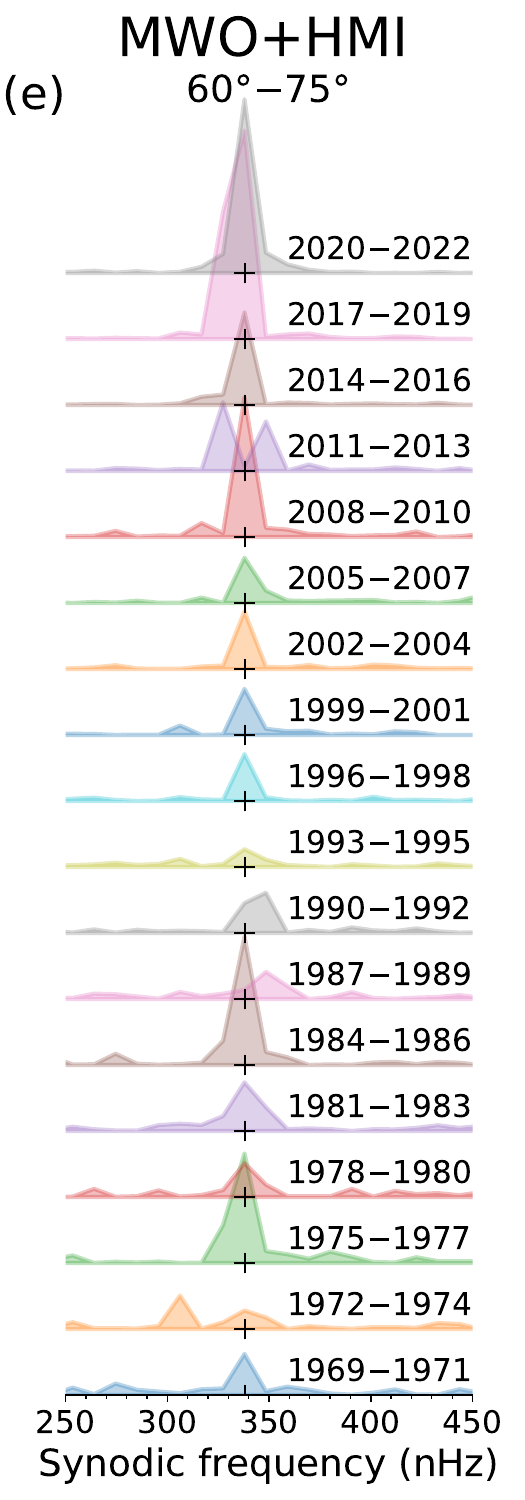}}
  \end{minipage}
  \tikzmark{start2}
  &
  \tikzmark{end2}
  \begin{minipage}[b]{64mm}
    \resizebox{\hsize}{!}{\includegraphics{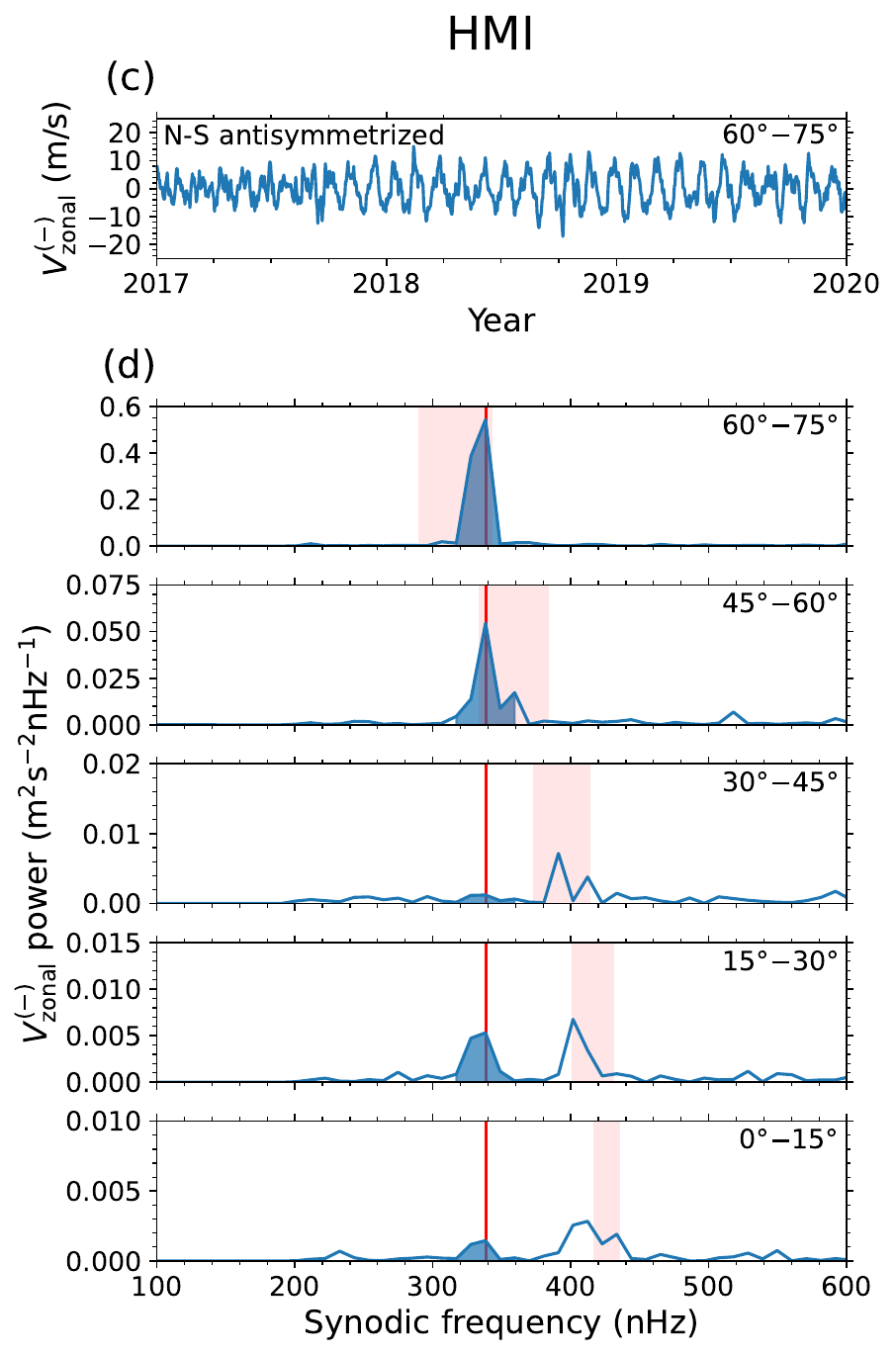}}
  \end{minipage}
  \end{tabular}

  \begin{tikzpicture}[overlay, remember picture]
  \draw[-stealth] ([xshift=0.3cm,yshift=3.0cm]pic cs:start1)
    -| ($([xshift=0.1cm,yshift=3.0cm]pic cs:start1)!0.5!([yshift=3.0cm]pic cs:end1)$) |-
  ([xshift=-0.2cm,yshift=6.4cm]pic cs:end1);
  \draw[-stealth] ([xshift=-0.2cm,yshift=7.7cm]pic cs:start2)
    -| ($([xshift=-0.1cm,yshift=6.4cm]pic cs:start2)!0.5!([yshift=6.4cm]pic cs:end2)$) |- 
  ([xshift=0.2cm,yshift=6.4cm]pic cs:end2);
  \end{tikzpicture}

  \caption{ \label{fig:pow-vs-nu} 
  Visibility of $\HL$ mode in consecutive three-year power spectra, using  MWO (1969\,--\,2010) and HMI (2011\,--\,2022) data.
Panel (a): Example MWO time series (1984--1986) of $\vzon^{(-)}$ averaged over latitudes $60\degr \leq \lambda < 75\degr$ (i.e., $\vband$).
Panel (b): Power spectra of $\vzon^{(-)}$ for the same time period in various latitudinal bands of  $15\degr$ (labeled at the top right corner of each panel). The vertical red lines denote the reference $\HL$ mode frequency $\nu_\HL^{\rm synodic}=338$~nHz  and the excess power around this frequency is highlighted in blue. The shaded pink areas indicate the differential rotation rate in the outer 5\% of the Sun from global helioseismology as viewed from the Earth.
Panels (c) and (d): Same as above but for an example HMI three-year time series during 2017--2019.
Panel (e): All available three-year power spectra of $\vband$ stacked on top of each other. The crosses highlight the reference frequency of 338~nHz. }
\end{figure*}

To compare with the Doppler data, we also reconstruct $\vlos^\HL$ using the $5\degr$-tile horizontal flow maps near the solar surface from the HMI ring-diagram pipeline \citep{Bogart2011a,Bogart2011b}. Because the ring-diagram flow maps ($\vec{V}_{\rm Carr}$) are given in the Carrington frame, it is necessary to apply at each position vector $\vec{r}$ (defined from the center of the Sun) the following velocity transformation from the Carrington frame to the Earth frame,
\begin{equation} \label{eq:carr2earth}
\vec{V}_\oplus = \vec{V}_{\rm Carr} + ({\Omega}_{\rm Carr} \vec{z}_\odot-{\Omega}_\oplus \vec{z}) \times  \vec{r} ,
\end{equation}
where $\Omega_{\rm Carr}/2\pi = 456$~nHz is the Carrington rotation rate, $\vec{z}_\odot$ is the unit vector that points along the solar rotation axis, $\Omega_\oplus/2\pi$ = (1~yr)$^{-1}$ = $31.7$~nHz, and $\vec{z}$ is the unit vector that points along the normal to the ecliptic plane. The second term on the right-hand side of Eq.~\eqref{eq:carr2earth}  does not affect the frequencies within the $\HL$ band-pass filter. Thus the $\vlos^{\HL}$ reconstructed from the ring-diagram flow maps is essentially $\vec{V}_{\rm Carr}$ projected onto the line of sight, followed by the $\HL$ mode filtering.

The middle column of Fig.~\ref{fig:cf-eigenfunc} shows the reconstructed $\vlos^\HL$ from the ring-diagram flow maps, to be compared with $\vlos^\HL$ maps from the direct Doppler.
The large-scale patterns and the velocity amplitudes are in very good agreement between the two data sets, especially at high latitudes.
This suggests in particular that the $\HL$ mode does not have any significant radial motion (since we did not consider this component in the reconstruction of the Doppler data). At lower latitudes, the large-scale features have much lower amplitudes but also agree.
A movie showing the evolution of $\vlos^\HL(x,y,t)$ and its ring-diagram reconstruction at a cadence of 1 day during 2010\,--\,2022 is available online \citep{movie}. At high latitudes, the agreement is good between the two data sets, although noise is higher in the ring-diagram data set. This is not too surprising as we expect helioseismic analysis to contain stochastic p-mode noise that is not present in Doppler data at a daily temporal cadence. 

Let us study the random noise in $\vlos^\HL$ in more detail.
We estimate the noise level by subtracting from $\vlos^\HL$ a spatial smoothing with a Gaussian FWHM of $15\degr$ and computing the rms of the residuals in time for each pixel.
As shown in Fig.~\ref{fig:cf-eigenfunc} (bottom row), the rms noise level in direct $\vlos^\HL$ is below $2$~m/s, with a minimum of $0.2$~m/s near disk center.
The noise level increases with disk radius because an angular distance of $5\degr$ corresponds to about 8 pixels near the disk center but only about one pixel near the limb.
The noise level in the reconstructed data from ring-diagram flow maps is also around $0.2$~m/s near disk center, but it increases much faster with disk radius than that in the direct Doppler data.
We also note that the reconstructed images from helioseismology are only available up to $98\%$ of the disk radius.
On the other hand, the Dopplergrams cover the entire solar disk, which is much useful to observe the high-latitude $\HL$ mode.

In the right column of Fig.~\ref{fig:cf-eigenfunc} we also compare the observations with the line-of-sight projection of the  theoretical velocity eigenfunction \citep{Gizon2021, Bekki2022a}. We find a reasonable agreement, however there are also noticeable differences between observations and theory.
In particular the tilt of the velocity pattern changes faster with latitude in the observations than in theory.

We remark that the tilt of the spiral pattern at high latitudes in Fig.~\ref{fig:vlos} is in the opposite direction to that of Fig.~\ref{fig:synoptic} since in the supersynoptic maps the corresponding longitude decreases toward the right along the horizontal axis. We also note that we avoided performing spatial Fourier transform, since we do not see the entire solar surface and LOS projection effects may complicate the interpretation of the spatial power spectra.

\section{Evolution of $\HL$ mode parameters}
\subsection{Power spectra from three-year data segments}

We break down the time series into three-year segments to extract the $\HL$ mode parameters as a function of time. We use the MWO data from 1969 to 2010 and the HMI data from 2011 to 2022, giving a total of 18 non-overlapping segments covering five solar cycles.

We compute the power spectra for each of these three-year segments. 
Above $45\degr$ latitude, the mode is detected in all data sets as narrow peaks of power near  synodic frequency 338~nHz well above noise level. This can be seen for two example time segments in Fig.~\ref{fig:pow-vs-nu} (panels b and d). Below $45\degr$, the mode is in the noise in the MWO data (Fig.~\ref{fig:pow-vs-nu}b), while it remains visible below  in the HMI power spectra (Fig.~\ref{fig:pow-vs-nu}d).
We note that there is also some excess power at frequencies corresponding to the local differential rotation (highlighted by the shaded pink areas), indicating that some velocity features are advected by the local rotation, such as flows around active regions \citep{gizon2004}, convective flows, or possibly critical-latitude modes \citep{Gizon2021, Fournier2022}.
The background power drops to zero below 200~nHz since we applied a high-pass filter to the time series of $\vzon$ to remove long-term trends including the `torsional oscillation' and $B_0$-angle variation.

The middle column of Fig.~\ref{fig:pow-vs-nu} shows the 3-year power spectra above $60^\circ$ latitude,
stacked on top of each other in the (synodic) frequency range 250\,--\,450~nHz.
The $\HL$ mode power is always present, although it is smaller during the periods 1972\,--\,1974 and 1993\,--\,1995.
We see that the noise level in the MWO data improved from 1982 when a new exit slit assembly was installed, and from 1986 when the fast scan program started, as reported by \citet{Ulrich1988}. We find that it is possible to follow the evolution of the mode amplitude over the full extent of the data sets (MWO, GONG, HMI) with a time resolution of at least three years.
It is remarkable that the MWO data enable us to add three decades of very useful observations.

\subsection{Estimation of mode parameters}

To improve the temporal sampling, we now consider sliding time windows of three years with central times separated by one year, such that each time window has an overlap of two years with the neighboring windows.
To extract the $\HL$ mode parameters in each time window, we choose to fit a Lorentzian function to the corresponding power spectra,
\begin{equation}
    L(\nu)= 
    \frac{A}{1 + [(\nu-\nu_\HL + 31.7 \textrm{nHz})/(w/2)]^2} + B,
\end{equation}
where $\nu_\HL$ is the sidereal mode frequency, $A$ is the mode power,  $w$ is the full width at half maximum of the Lorentzian,  and $B$ the constant background over the fitting range 200\,--\,500~nHz.
The total power of the $\HL$ mode under the Lorentzian profile is  $\pi Aw$.
The rms and maximum values of the mode velocity amplitude ($\vband$) are given by  $\sqrt{\pi Aw}$ and $\sqrt{2\pi Aw}$ respectively.

We note that we padded each 3-yr time segment of $\vband$ with zeros to have a length of $N_t=8192$ and a frequency resolution of $\Delta\nu=1.4$~nHz.
The original frequency resolution from a 3-yr time segment ($10.6$~nHz) is coarser than the line width of the $\HL$ mode (4\,--\,6~nHz; see \S~\ref{sect:fullres}), making it tricky to fit. The mode power $\pi Aw$ is unaffected by the zero padding as the duty cycle  $\eta$  has taken this into account in Eq.~\eqref{eq:power}; however, the line profile is broadened by the padding so we do not consider $w$ and $A$ individually. We also note that the $\vzon$ defined in Eq.~\eqref{eq:vzon} is smaller than the corresponding longitudinal velocity $v_\phi$. Using synthetic data, we find that the ratio of $v_\phi$ to $\vzon$ ranges from $2.1$ to $3.3$, depending on the latitude and the model used. We choose to scale the mode amplitude measured from $\vband$ by a factor of $2.8$ so that it has the same magnitude as the one measured from the $v_\phi$ component of HMI ring-diagram flow maps during solar minimum of cycle 24/25.

\begin{figure}
    \centering
    \resizebox{\hsize}{!}{\includegraphics{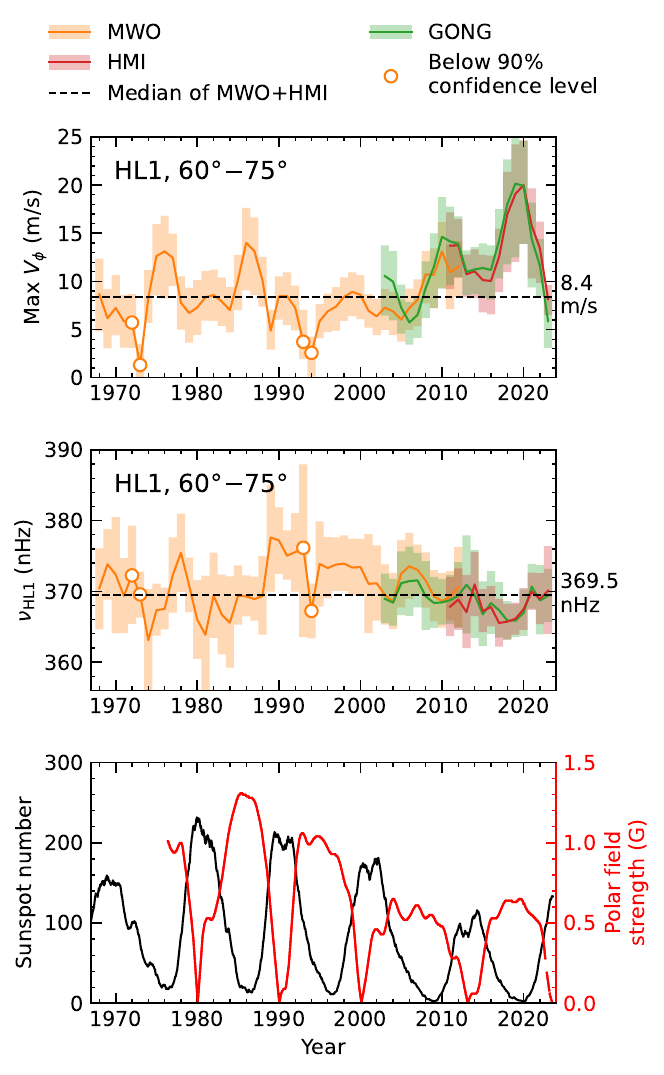}}
    \caption{ \label{fig:amp-vs-t}
    Temporal variations of the amplitude (\emph{top panel}) and frequency (\emph{middle panel}) of the $\HL$ mode, together with the sunspot number and the polar field strength (\emph{bottom panel}).
    The orange, green, and red solid curves show the results from MWO, GONG and HMI data sets, respectively, with the shaded areas showing the 68\% confidence interval estimated from 10\,000 Monte Carlo simulations.
    The open circles indicate that the excess power around the mode frequency is below 90\% confidence level.
    The horizontal dashed lines represent the median value over MWO (1968\,--\,2010) and HMI (2011\,--\,2023) data sets.
    }
\end{figure}

\begin{figure}
    \centering
    \includegraphics[width=\hsize]{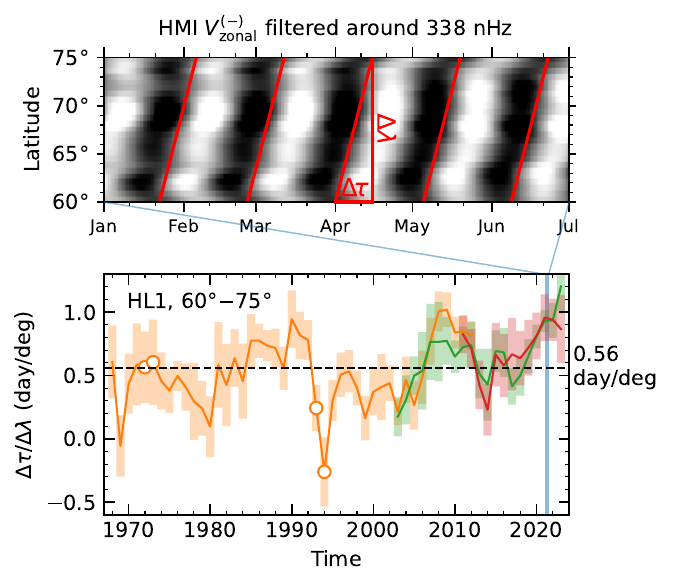}
    \caption{ \label{fig:tilt}
    Temporal variations of $\HL$ mode velocity pattern.
    Top panel: HMI $\vzon^{(-)}$ in the first half of 2021 applied with a band-pass filter around $\nu_\HL^{\rm synodic}$.
    The slope of the striped pattern in $\vzon^{(-)}$ is highlighted by the tilted red lines.
    The $\Delta\tau$ (time lag) and $\Delta\lambda$ (latitudinal separation) that determine the slope of the striped pattern $\Delta\tau/\Delta\lambda$ are indicated in the plot (the two legs of the red right triangle).
  Bottom panel: 
    $\Delta\tau/\Delta\lambda$ as a function of time,    
    using the same color conventions as in Fig.~\ref{fig:amp-vs-t}.
    }
\end{figure}

\subsection{Evolution of mode amplitude and frequency}

Figure~\ref{fig:amp-vs-t} shows the temporal variations of the measured amplitude and frequency of the $\HL$ mode. The results from the three data sets during the overlapping periods are consistent to a large extent.
The mode amplitude is mostly above 5~m/s with a median value of $8.4$~m/s whereas the mode frequency has a median value of $369.5$~nHz with a peak-to-peak variation of about 10~nHz.
The mode amplitude reaches its overall maximum during the solar minimum of cycle 24/25 right after the weakest cycle 24 in the past five cycles. The 11-yr variation can be seen in the mode amplitude, especially during cycles 23 and 24. On the other hand, there is no clear 11-yr variation in the mode frequency.

\begin{figure*}
    \centering
    \begin{tabular}{cc}
    \begin{minipage}[b]{9cm}
        \resizebox{\hsize}{!}{\includegraphics{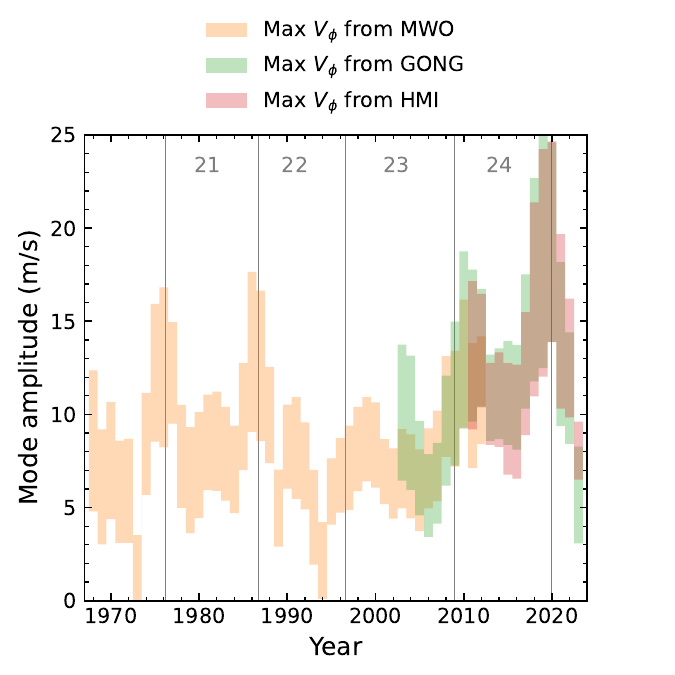}}
    \end{minipage}
    &
    \begin{minipage}[b]{9cm}
        \resizebox{\hsize}{!}{\includegraphics{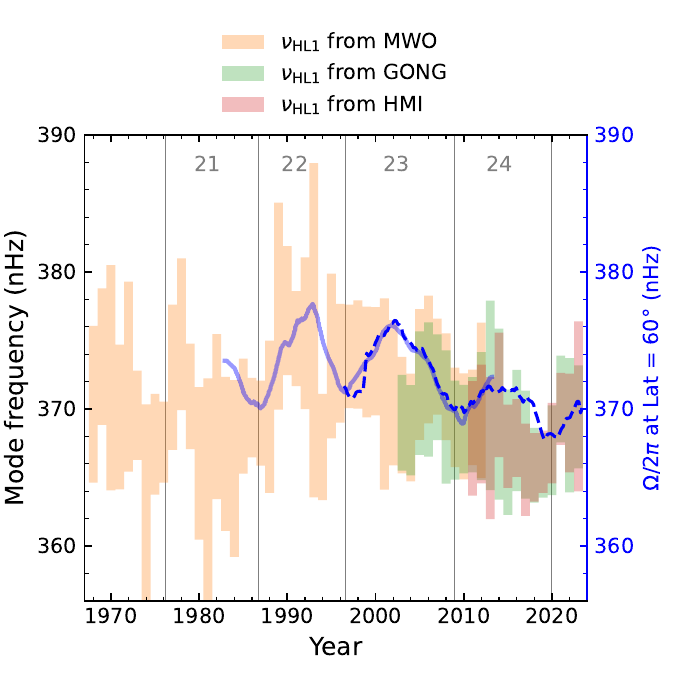}}
    \end{minipage}
    \end{tabular}
    \caption{ \label{fig:mode-vs-rot}
    Temporal variation of the mode amplitude (\emph{left}) and frequency (\emph{right}) taken from Fig.~\ref{fig:amp-vs-t}, together with surface rotation rates at $60\degr$ latitude obtained from MWO Doppler velocity (1983\,--\,2013) in solid light blue, and from global helioseismology using MDI (1996\,--\,2010) and HMI (2010\,--\,2023) data in dashed blue.
    The beginnings of cycles 21\,--\,25 (solar minima) are marked with vertical lines and the cycle numbers are denoted at the top.
     }
\end{figure*}

\subsection{Evolution of mode velocity pattern}

In addition to the mode amplitude and frequency, we also measure the slope of the striped pattern seen in the supersynoptic maps of $\vzon$ at high latitudes.
We define the slope of the striped pattern as the ratio of $\Delta\tau/\Delta\lambda$, where $\Delta\tau$ is the time lag between the time series of $\vzon$ at different latitudes and $\Delta\lambda$ is the latitudinal separation thereof, as depicted in the top panel of Fig.~\ref{fig:tilt}.
In practice we determine the time lag that maximizes the correlation between $\vzon$ at adjacent latitudes and average over latitude weighted by a raised cosine between latitudes $60\degr$ and $75\degr$.
The bottom panel of Fig.~\ref{fig:tilt} shows its temporal variation.
The measured $\Delta\tau/\Delta\lambda$ varies mostly between $0.2$ and 1~day/deg with a medium value of $0.56$~day/deg.
Although there are two points below zero (years 1969 and 1994), they fall within the range of uncertainty.
It seems fair to say that $\Delta\tau/\Delta\lambda$ does not change sign.
The $\Delta\tau/\Delta\lambda$ seems to have larger values during solar minima and exhibits an 11-yr variation to some extent, though not as clearly as in the mode amplitude.
We note that the time lag $\Delta\tau$ has a dependence on the mode frequency.
Nonetheless, the variation in the measured $\nu_\HL$ (about $\pm5$~nHz) only affects $\Delta\tau/\Delta\lambda$ by $\pm1.5\%$ which is negligible.

\subsection{Correlations between mode parameters,  sunspot number, and  polar field}

In order to examine which periodicities that are present in the time series of the $\HL$ mode parameters, we combine the measurements from MWO (1968\,--\,2010) and HMI (2011\,--\,2022) to form a 55-yr long data set, and we compute a temporal power spectrum for each of the parameters (velocity amplitude, frequency, and eigenfunction tilt).
The right column of Fig.~\ref{fig:11periodicity} shows these three power spectra.
We see that there is a clear 11-yr periodicity in the mode velocity amplitude, while the power spectrum for the mode frequency does not show a distinct peak at (11~yr)$^{-1}$ = $2.88$ nHz.
The second and third plots on the left column of Fig.~\ref{fig:11periodicity} highlight (in red) the 11-yr component of the evolution of the velocity amplitude and the tilt. We see that they are anti-correlated with the sunspot number.
The correlation coefficients between the mode parameters and the sunspot number are $-0.5$ for the mode amplitude and  $-0.34$ for the eigenfunction tilt.
Since there are only 18 independent points in a period of 55 years, we find that coefficients lower than $0.47$ in absolute value are not significant ($p > 0.05$).
Thus only the anticorrelation between the mode amplitude and the sunspot number is significant.

We also compute the correlation coefficients between the mode parameters and the polar field strength over the period 1977--2022 shown in the bottom panel of Fig.~\ref{fig:amp-vs-t}.
Surprisingly, these correlation coefficients are all below $0.1$ in absolute value, which means there is no correlation between the mode parameters and the polar field strength.

\subsection{Correlations between mode parameters and  rotation rate}

We now compare the mode parameters with the solar surface rotation rate at $60\degr$ latitude where the local rotation rate is close to the mode frequency.
From 1996, we use the solar rotation rate obtained from global helioseismology \citep{Larson2015,Larson2018}.
The data series used are \texttt{mdi.vw\_V\_sht\_2drls} from 1996 to 2010 and \texttt{hmi.vw\_V\_sht\_2drls} from 2010 to 2022, at intervals of 72 days with \texttt{NACOEFF}=6.
The rotation rate is smoothed with a 360-day running window.
We then  extend the measurements back to 1983 using the MWO Doppler velocity data, as processed by \citet[, their table~1 and figure~4]{Ulrich2023}.  The MWO rotation rate  is shifted by an offset of 16 nHz so it matches the near-surface rotation rate from global helioseismology during the overlapping time period. 
Also, because the global helioseismology data are north-south symmetric by construction, we symmetrize the MWO data.
We remark that the surface rotation rate from the MWO Doppler velocity  starts from 1983, despite the fact that the MWO Doppler data have been available since 1967. This is because there exists a systematic error that offsets the Doppler signals prior to 1983 \citep[, section~2]{Ulrich2023}.
Since we only consider the signals in the inertial frequency range, this systematic error may not bias the measured mode parameters.

The right panel of Fig.~\ref{fig:mode-vs-rot} compares the evolution of the mode frequency and the surface rotation rate at $60\degr$ latitude for 1983\,--\,2023, together with the mode frequency  (the shaded region covers the error range of $\pm 1 \sigma$).
Both curves show the same long-term decrease, which amounts to about $-7$ nHz over 30 years. This decrease has been reported before at this latitude during cycles 23 and 24 throughout the convection zone \citep[, their figure 2]{Basu2019}. 
In addition, the rotation rate also shows an 11-year component at the level of $\pm 2.5$~nHz associated with the high-latitude branch of the so called `torsional oscillation' \citep[e.g.,][]{Howe2009,Howe2013,Basu2016}. These small variations in the rotation rate are within the error bars of the inertial mode frequency ($\pm 3$~nHz or so, for a three year time window). The mode frequency measurements are too noisy to reveal any variation at this level of precision. 
Overall, the  correlation coefficient between the mode frequency and the rotation rate at $60\degr$ is $0.55$. This correlation coefficient is significant ($p=0.05$).

The left panel of Fig.~\ref{fig:mode-vs-rot} compares the evolution of the mode amplitude with the timing of the solar cycles.
The times of solar minima are indicated on the plot with vertical bars. We find that the times at which the mode amplitude reaches its maximum coincide with the beginnings of cycles 21, 22, and 25.
But when the mode amplitude is smaller (solar minima of cycles 22/23 and 23/24), the mode has its maximum amplitude 1\,--\,3 years after the solar minima.
Furthermore, cycle 21 is the strongest sunspot cycle in the past five decades and the polar field strength in the subsequent solar minimum is notably strong; however, the mode amplitude is not particularly small in that period.
We see that the relative mode amplitude variations are of order unity, while the relative change in rotation due to the `torsional oscillation' is about 1\%.
We find a correlation coefficient between the mode amplitude and the rotation rate at $60\degr$ of $-0.82$. This anticorrelation  is very significant ($p=0.0006$).

\section{Summary and discussion}

We have computed synoptic maps of zonal velocity $\vzon$ using LOS Doppler velocity from MWO, GONG, and HMI, covering a total of 57 years.
We have measured the $\HL$ mode parameters from these synoptic maps at high latitudes.
The time series of mode amplitude and eigenfunction tilt exhibit an 11-yr periodicity, whereas the 11-yr periodicity in mode frequency is below noise level.
The time series of mode amplitude and eigenfunction tilt are weakly anticorrelated with the sunspot number.
However, the anticorrelation with the high-latitude rotation rate is more pronounced, particularly for the mode amplitude.
The mode amplitude peaks coincide with solar minima when the mode amplitude is large, but lag by 1\,--\,3 years when the mode amplitude is smaller.
The long-term trend of the mode frequency follows the rotation rate near the critical latitude.
The results presented in this work contain a wealth of information about the $\HL$ mode.
Improved  linear and nonlinear models of inertial modes (with magnetic field) are needed to interpret the observations and further constrain the physics of the solar interior.

The temporal changes in  mode frequency are very small and can probably be explained using the linearized wave equation together with a prescription of the variations in solar  rotation and solar structure. In particular, the solar-cycle variations of rotation (`torsional oscillation') are known from p-mode helioseismology throughout the convection zone, and their effect on the $\HL$ mode frequency can be inferred using a linear eigenvalue solver \citep{Bekki2022a} or  a first-order perturbation method \citep[see work by][ for the equatorial Rossby modes]{Goddard2020}. However, it remains to be seen if changes in the rotation alone will be sufficient to explain the data. The changes in the background entropy and the direct effect of the magnetic field on the mode may have some importance as well. In turn, the measurements of mode frequency   present an opportunity to constrain these changes in the convection zone (unlike p modes,  inertial modes have large kinetic energy density deep in the convection zone).

The much larger variations  in  mode amplitude and mode eigenfunctions will require more sophisticated modeling tools. 
Recent non-linear 3D hydrodynamic simulations carried out by \citet{Bekki2024} have revealed the complex relationship between the  baroclinically-unstable high-latitude modes, the entropy gradient, and the rotation rate. It was shown that the modes may transport entropy toward the equator and thus back react on the latitudinal differential rotation.
Such work represents a first step in the understanding of the  dynamics of the $\HL$ mode, but 
will have to be extended to account for the effects of the dynamo.

\begin{acknowledgements}
We are very grateful to Roger K.~Ulrich for providing the data from the synoptic program at the
150-Foot Solar Tower of the Mt. Wilson Observatory. The
Mt. Wilson 150-Foot Solar Tower is operated by UCLA, with
funding from NASA, ONR and NSF, under agreement with the
Mt. Wilson Institute. 
The HMI data used are courtesy of NASA/SDO and the HMI science team.
SOHO is a project of international cooperation between ESA and NASA.
This work utilizes GONG data from the National Solar Observatory (NSO), which is operated by AURA under a cooperative agreement with NSF and with additional financial support from NOAA, NASA, and USAF.
The sunspot numbers are from the World Data Center SILSO, Royal Observatory of Belgium, Brussels.
The polar magnetic ﬁeld data are from the Wilcox Solar Observatory (http://wso.stanford.edu/Polar.html), courtesy of J.T. Hoeksema. The Wilcox Solar Observatory is currently supported by NASA.
The data were processed at the German Data Center for SDO (GDC-SDO), funded by the German Aerospace Center (DLR).
L.G. acknowledges support from the NYU Abu Dhabi Center for Space Science.
\end{acknowledgements}


\bibliographystyle{aa}
\bibliography{ref}

\clearpage
\begin{appendix}

\section{Additional figures}

\begin{figure*}
  \centering
  \includegraphics[width=14cm]{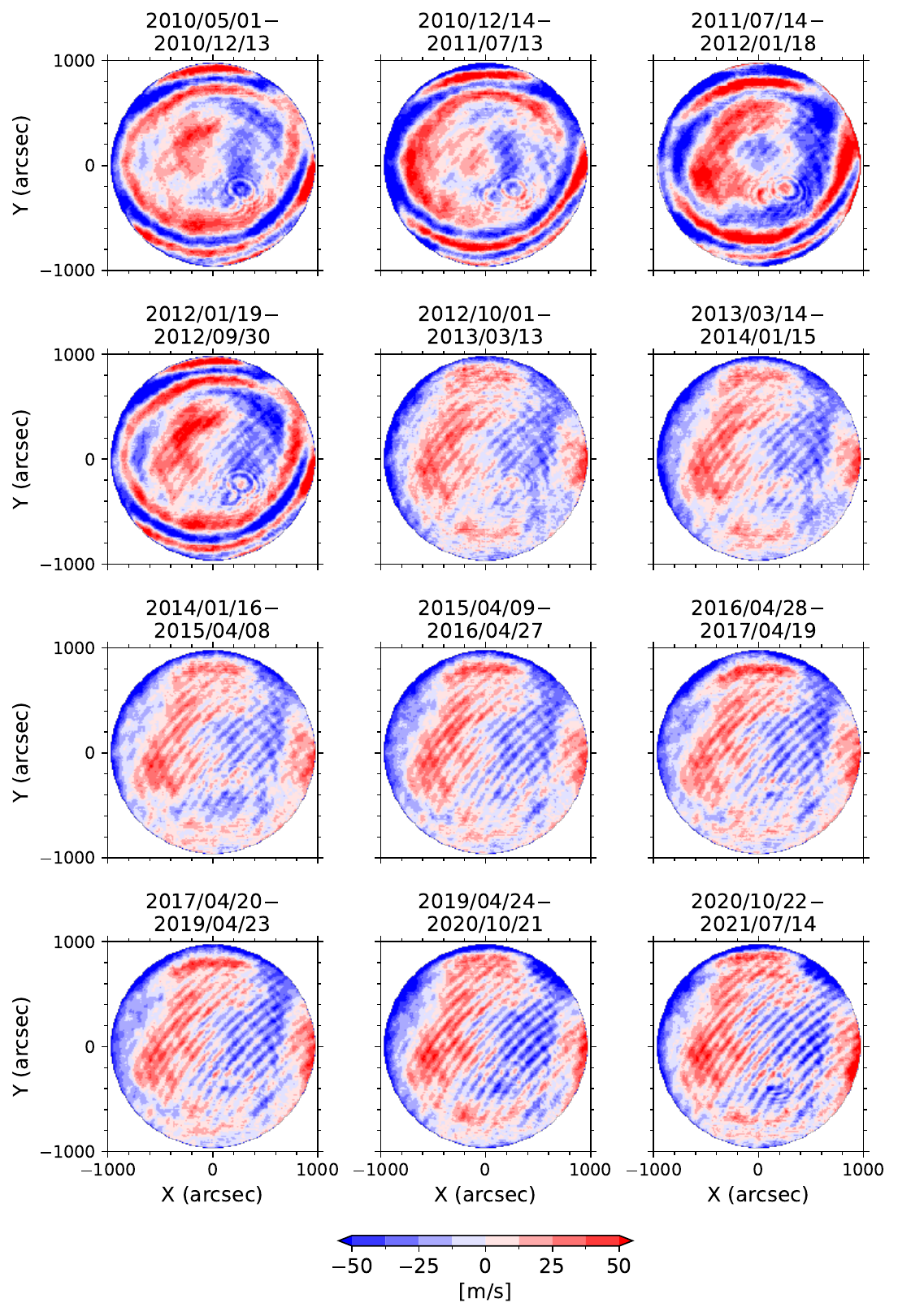}
  \caption{ \label{fig:fringes}
  Median values of the Dopplergrams over the time periods between the HMI retuning dates listed at http://jsoc.stanford.edu/doc/data/hmi/hmi\_retuning.txt. 
  The images have been remapped to the same resolution (10\arcsec per pixel), with the disk center aligned north-up, and the $\vb$ component subtracted.
  The color scale is saturated at $\pm50$~m/s.
  \citet[][, their figure 2]{Hathaway2015} shows a similar pattern in the HMI data from 2012 to 2013.
  The interference fringes appear as wavy patterns in the HMI images, caused by imperfections in the tunable-filter elements of the optical system. These unwanted patterns in the Dopplergrams exceed 50 m/s in magnitude, which is significantly higher than the amplitude of the $\HL$ mode. 
  The large-scale fringes were substantially removed by the HMI team from 1 October 2012 \citep{Couvidat2016}, which explains why the fringes in the last eight panels are much weaker than  in the first four panels. While the temporal frequency of these patterns is very low and far from the $\HL$ mode frequency, the patterns shift during HMI instrument re-tuning, causing jumps in the Doppler time series and increasing the background power in the Fourier domain. Removing these fringes enhances the visibility of  the $\HL$ mode in the synoptic maps of $\vzon$ and improves the signal-to-noise ratio of the mode amplitude measurement.
  }
\end{figure*}
\clearpage

\begin{figure*}
  \centering
  \includegraphics[width=17cm]{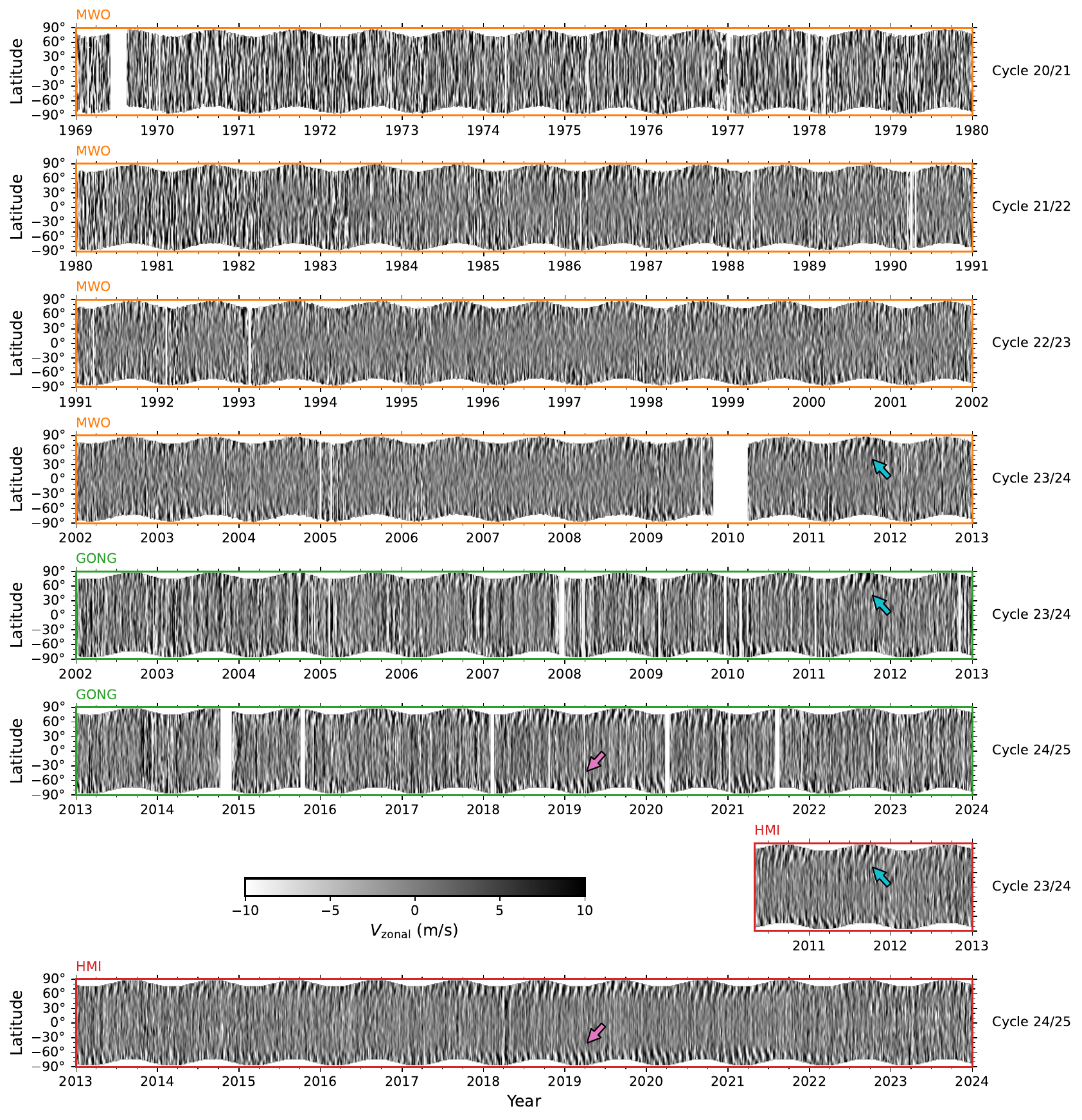}
  \caption{ \label{fig:synoptic-all}
  Synoptic maps of $\vzon$ as  functions of latitude and time, obtained from direct $\vlos$ of HMI, GONG and MWO observations.
  For clarity, the maps were Gaussian smoothed in latitude with a FWHM of $10\degr$.
  The arrows with the same color highlight the stripes for different data sets in the same time period.
  }
\end{figure*}
\clearpage

\begin{figure*}
  \centering
  \includegraphics[width=17cm]{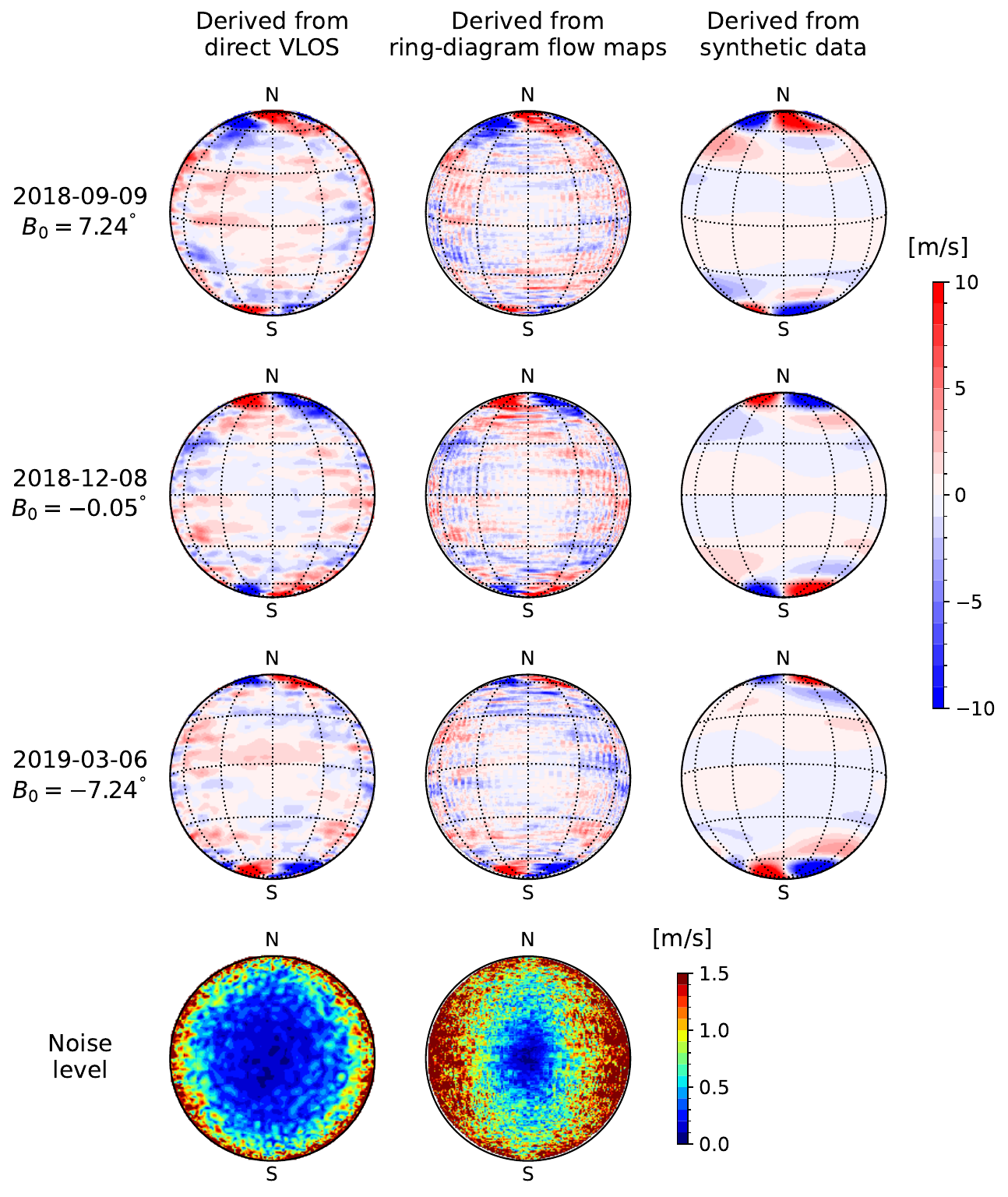}
  \caption{ \label{fig:cf-eigenfunc}
  Comparison of the $m=1$ high-latitude mode oscillations in the line-of-sight Doppler velocity as viewed from the Earth, obtained using various methods, when the solar tilt angle $B_0$ is $7.24\degr$ (top), $0\degr$ (middle), and $-7.24\degr$ (bottom).
  \emph{Left column}: Observed Doppler velocity $\vlos$ from HMI data band-pass filtered around $\nu_\HL^{\rm synodic}$ ($\vlos^\HL$, see Eq.~\eqref{eq:filt}).
  For a better comparison with the middle column, the images were Gaussian smoothed with a FWHM of $5\degr$.
  \emph{Middle column}: reconstructed $\vlos$, obtained by projecting the $5\degr$-tile horizontal flow maps from the HMI ring-diagram analysis onto the line of sight and applying the same filter as in the first column.
  \emph{Right column}: Line of sight projection of mode eigenfunction obtained using a linear eigenvalue solver \citep{Bekki2022a},  scaled to an amplitude comparable to the observations.
  \emph{Bottom row}: Maps of noise in $\vlos^\HL$ estimated from the temporal variance of the data at each spatial point.
  }
\end{figure*}
\clearpage

\begin{figure*}
  \centering
  \includegraphics[width=18cm]{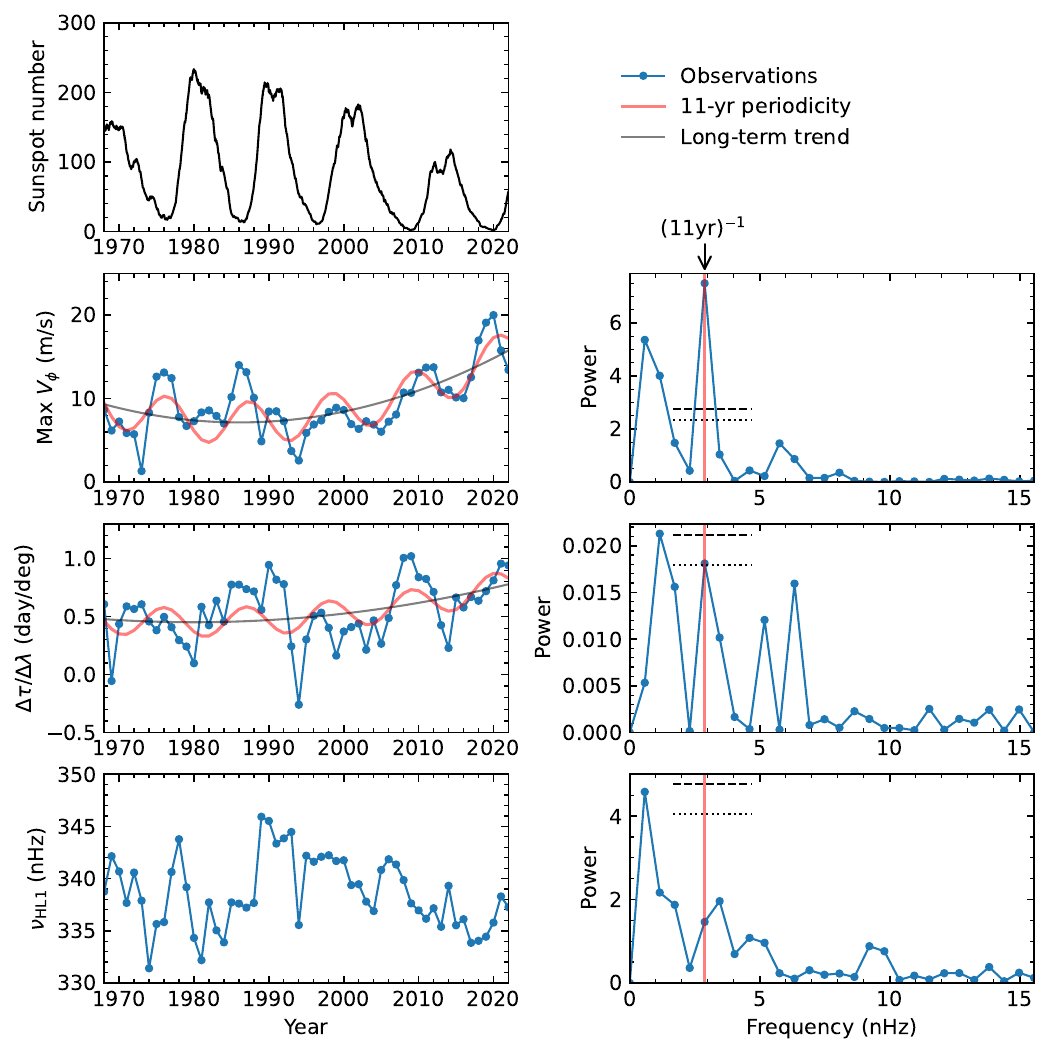}
  \caption{ \label{fig:11periodicity}
  Eleven-year periodic components and power spectra of $\HL$ mode parameters. 
  \emph{Left column, from top to bottom}:  sunspot number,  $\HL$ mode amplitude, eigenfunction tilt, and  frequency.
  The 55-yr long time series   combine the MWO (1968\,--\,2010) and HMI (2011\,--\,2022) measurements taken from Figs.~\ref{fig:amp-vs-t} and \ref{fig:tilt}.
  The gray curves are fitted parabolas representing the long-term trend.
  The pink curves show the 11-yr periodicity in the data added to the long-term trend.
  \emph{Right column}: power spectra of the time series from the left column with the temporal mean subtracted.
  The vertical pink lines indicate the frequency of $(11\textrm{ yr})^{-1} = 2.88$~nHz that corresponds to the average sunspot cycle period. The horizontal dashed and dotted lines represent the 95\% and 90\% confidence levels, respectively. The 11-year periodicity is highly significant for the mode amplitude, and only marginally significant for the eigenfunction tilt. No evidence is found for an 11-year periodicity in the mode frequency.
  }
\end{figure*}

\end{appendix}

\end{document}